\documentclass[%
 reprint,
superscriptaddress,
 amsmath,amssymb,
 aps,
prb,
]{revtex4-1}

\usepackage{graphicx}
\usepackage{dcolumn}
\usepackage{bm}

\begin{document}

\preprint{APS/123-QED}

\title{Ferroelectric Negative Capacitance Domain Dynamics}

\author{Michael Hoffmann}
\email{michael.hoffmann@namlab.com}
\affiliation{Department of Electrical Engineering and Computer Sciences, University of California, Berkeley, CA 94720, USA}
\affiliation{NaMLab gGmbH, Noethnitzer Str. 64, D-01187, Dresden, Germany}

\author{Asif Islam Khan}
\affiliation{School  of Electrical  and Computer  Engineering, Georgia Institute of Technology, Atlanta, GA 30332, USA }

\author{Claudy Serrao}
\affiliation{Department of Electrical Engineering and Computer Sciences, University of California, Berkeley, CA 94720, USA}

\author{Zhongyuan Lu}
\affiliation{Department of Electrical Engineering and Computer Sciences, University of California, Berkeley, CA 94720, USA}
 
\author{Sayeef Salahuddin}
\affiliation{Department of Electrical Engineering and Computer Sciences, University of California, Berkeley, CA 94720, USA}
\affiliation{Material Science Division, Lawrence Berkeley National Laboratory, Berkeley, CA 94720, USA}


\author{Milan Pe\v{s}i\'{c}}
\affiliation{NaMLab gGmbH, Noethnitzer Str. 64, D-01187, Dresden, Germany}

\author{Stefan Slesazeck}
\affiliation{NaMLab gGmbH, Noethnitzer Str. 64, D-01187, Dresden, Germany}

\author{Uwe Schroeder}
\affiliation{NaMLab gGmbH, Noethnitzer Str. 64, D-01187, Dresden, Germany}

\author{Thomas Mikolajick}
\affiliation{NaMLab gGmbH, Noethnitzer Str. 64, D-01187, Dresden, Germany}
\affiliation{Chair of Nanoelectronic Materials, TU Dresden, D-01062 Dresden, Germany}


\begin{abstract}
Transient negative capacitance effects in epitaxial ferroelectric Pb(Zr$_{0.2}$Ti$_{0.8}$)O$_3$ capacitors are investigated with a focus on the dynamical switching behavior governed by domain nucleation and growth. Voltage pulses are applied to a series connection of the ferroelectric capacitor and a resistor to directly measure the ferroelectric negative capacitance during switching. A time-dependent Ginzburg-Landau approach is used to investigate the underlying domain dynamics. The transient negative capacitance is shown to originate from reverse domain nucleation and unrestricted domain growth. However, with the onset of domain coalescence, the capacitance becomes positive again. The persistence of the negative capacitance state is therefore limited by the speed of domain wall motion. By changing the applied electric field, capacitor area or external resistance, this domain wall velocity can be varied predictably over several orders of magnitude. Additionally, detailed insights into the intrinsic material properties of the ferroelectric are obtainable through these measurements. A new method for reliable extraction of the average negative capacitance of the ferroelectric is presented. Furthermore, a simple analytical model is developed, which accurately describes the negative capacitance transient time as a function of the material properties and the experimental boundary conditions. 
\end{abstract}

\maketitle


\section{Introduction}

To further continue the remarkable trend of miniaturization of nanoelectronic devices in integrated circuits, fundamental power density constraints have to be overcome in the near future.\cite{theis_its_2010} These physical limits stem from the thermal broadening of Fermi-Dirac statistics in any device using the injection of electrons over an energy barrier as the operation principle. Therefore, the ultimate lower limit of how much voltage is needed to change the current in such devices by one order of magnitude is given by ln(10)$k_B T/q$, where $k_B$ is the Boltzmann constant, $T$ is the temperature and $q$ is the elementary charge.\cite{zhirnov_negative_2008}

In 2008, Salahuddin and Datta proposed to use negative capacitance (NC) in a ferroelectric material to overcome this sometimes called "Boltzmann tyranny",\cite{zhirnov_negative_2008} by using the ferroelectric as an insulating layer between the controlling gate electrode and the semiconductor channel.\cite{salahuddin_use_2008} This new physical property of NC was predicted from Landau-Devonshire theory of ferroelectric phase transitions and was already indirectly confirmed in several different ferroelectric-dielectric heterostructures and superlattices.\cite{islam_khan_experimental_2011,gao_room-temperature_2014,appleby_experimental_2014,zubko_negative_2016,kim_time-dependent_2016} Recently, first direct measurements of NC in ferroelectric capacitors in series with a resistor have been reported,\cite{khan_negative_2014,hoffmann_direct_2016} opening up new ways to characterize and investigate this novel physical phenomenon, which is to some degree still controversially discussed in the literature.\cite{cano_multidomain_2010,krowne_examination_2011,song_alternative_2016} This cautious attitude towards the topic is comprehensible, since there is a considerable lack of understanding of the physics of NC in general. This is especially troubling, considering first NC devices overcoming the Boltzmann limit have been reported recently.\cite{lee_prospects_2015,li_sub-60mv-swing_2015,lee_physical_2016,zhou_ferroelectric_2016} To further improve these promising initial results, a more detailed understanding of the underlying physics and more reliable modeling approaches will be crucial.

\begin{figure}
\includegraphics[width=6.5cm]{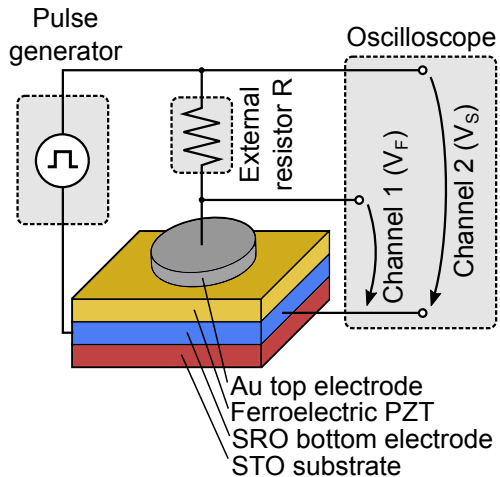}
\caption{Schematic sample structure and measurement circuit applied to directly measure negative capacitance voltage transients via an oscilloscope.} \label{fig1}
\end{figure}

While almost all modeling efforts of NC up to now have applied homogeneous single-domain Landau-Khalatnikov theory, the more subtle but critical influence of domain formation has only been discussed in a very limited manner so far. \cite{cano_multidomain_2010,zubko_negative_2016,Hoffmann_Modeling_2017} However, first reports on domain dynamics in ferroelectric capacitors during switching are beginning to emerge.\cite{yuan_switching-speed_2016,hoffmann_direct_2016,zhu_negative_2017} Understanding the influence of ferroelectric domain formation and dynamics on NC effects will be critical for the utilization in actual devices. NC voltage transients were measured on high quality epitaxially grown ferroelectric lead zirconate titanate (PZT) capacitors connected in series to a resistor as shown in the schematic in Fig. \ref{fig1}. It will be elucidated how ferroelectric material properties like negative capacitance per area, activation field and internal loss can be extracted from such measurements using a newly developed analytical model.

\section{Experimental Setup}

Ferroelectric thin films of Pb(Zr\textsubscript{0.2}Ti\textsubscript{0.8})O\textsubscript{3} (PZT) were grown epitaxially using pulsed laser deposition on metallic SrRuO\textsubscript{3} (SRO) buffered SrTiO\textsubscript{3} (001) substrates. The 100 nm PZT and the 20 nm SRO films were deposited at 720 $^{\circ}$C and 630 $^{\circ}$C, respectively. During the growth, the oxygen partial pressure was kept at 100 mTorr and afterwards the heterostructure
was slowly cooled down at 1 atm of oxygen partial pressure and a rate of \mbox{-5 $^{\circ}$C/min} to room temperature. Laser pulses of \mbox{100 mJ} of energy and $\sim$4 mm\textsuperscript{2} of spot size were used to ablate the targets. Ti/Au top electrodes were ex-situ deposited by e-beam evaporation and then patterned
using standard lithographic techniques into square dots an area of \mbox{(10 $\mu$m)\textsuperscript{2}} to \mbox{(50 $\mu$m)\textsuperscript{2}}. The SRO layer was used as the bottom electrode.

The schematic measurement setup is shown in Fig. \ref{fig1}. An external resistor ($R = 200 \ \Omega\text{ to }200 \ \text{k}\Omega$) is connected to the top electrode of the ferroelectric capacitor. An Agilent 81150A pulse function generator was used to apply square voltage pulses (with rise and fall times of 50 ns) to the series connection of the ferroelectric capacitor and the resistor. The voltages of the pulse generator ($V_S$) and across the ferroelectric ($V_F$) were directly probed with a digital oscilloscope (GW Instek GDS-3354). The parasitic parallel capacitance $C_p = 35$ pF of the measurement setup was determined by measuring the charging time constant $\tau = RC_p$, with the ferroelectric capacitor disconnected. All measurements were carried out at room temperature. 

\section{Modeling of Ferroelectric Domain Dynamics}

\subsection{Time-dependent Ginzburg-Landau Theory of Ferroelectric Phase Transitions}

According to the Landau-Ginzburg-Devonshire theory of ferroelectric phase transitions,\cite{cao_theory_1991} the Gibbs free energy per volume $u$ of a ferroelectric material can be expanded in terms of an order parameter, which is the electric polarization $P$, as

\begin{equation}
u = \alpha P^2 + \beta P^4 + \gamma P^6 - EP + k(\nabla P)^2,
\label{eq1}
\end{equation}

where $\alpha$, $\beta$ and $\gamma$ are the ferroelectric anisotropy constants, $E$ is the electric field and $k$ is the domain coupling constant, which is related to spatial non-uniformity of $P$. While the $-EP$ term in Eq. (\ref{eq1}) describes the electrostatic energy, the Ginzburg term $k(\nabla P)^2$ accounts for the energy penalty for spatial variations of $P$ and therefore also includes domain wall energy contributions. In ferroelectrics, $\alpha$ is always negative below the Curie temperature $T_C$, which is the origin of the NC behavior in this class of materials. For second-order phase transitions, $\beta > 0$ and $\gamma = 0$. On the other hand, for the more common case of first-order phase transitions, $\beta < 0$ and $\gamma > 0$.

When the dynamical evolution of $P$ is of interest, the most straight forward approach is to apply the Landau-Khalatnikov equation \cite{landau_anomalous_1954}

\begin{equation}
-\frac{\partial u}{\partial P} = \rho \frac{\partial P}{\partial t},
\label{eq2}
\end{equation}

where $\rho$ is a damping parameter of unit [$\Omega $m], which is a measure of the loss in the material. It should be noted that Eq. (\ref{eq2}) does not describe the intrinsic dynamics during ferroelectric switching, but is only a diffusive limit approximation.\cite{chatterjee_intrinsic_2017} In this limit, the dynamics of the system is described by the domain wall velocity, which depends on the parameter $\rho$. For investigations beyond the diffusive limit approximation of  Eq. (\ref{eq2}), molecular dynamics simulations would have to be carried out. However, this is beyond the scope of this work.

Eq. (\ref{eq1}) in conjunction with Eq. (\ref{eq2}) is essentially identical to the time-dependent Ginzburg-Landau formulation which is known from the theory of superconductivity \cite{ginzburg_theory_1950} and is commonly used in ferroelectric phase field modeling.\cite{wang_phase-field_2004,chen_phase_2008} In the following, the numerical implementation of the time-dependent Ginzburg-Landau model used for simulation of ferroelectric NC voltage transients is described.

\subsection{Numerical Implementation}

In order to keep the model as simple as possible without losing its descriptive capabilities, a few assumptions are made based on the experimental boundary conditions. Since ferroelectric capacitors of 100 nm thickness with metallic electrodes are investigated, depolarization fields due to imperfect screening at the electrode interfaces should play no significant role and are therefore neglected here. Furthermore, only the out-of-plane ($z$-axis) contributions to the polarization of the ferroelectric are considered, which is reasonable since the measured PZT films are c-axis oriented and switched through the application of rather high electric fields ($E \ge 0.4$ MV/cm). $P$ can vary in the $x$-$y$ plane of the capacitor. Therefore, only 180$^\circ$ domain walls can be modeled. While ferroelastic 90$^\circ$ domain wall are known to exist in PZT thin films,\cite{wang_phase-field_2004,nagarajan_dynamics_2003} their effect on transient NC seems to be negligible in our experiments, since films of 100 nm thickness and below show almost no 90$^\circ$ domain walls due to strain imposed by the substrate.\cite{li_investigation_2007} Other effects related to strain, e.g. polarization-strain coupling are not considered in our model, see Eq. (\ref{eq1}).

Spatial derivatives of $P$ are calculated by a finite difference method on a uniform square grid in the $x$-$y$ plane. The spatial difference between adjacent points in $x$- and $y$-direction are identical, i.e. $\Delta x = \Delta y$. Each point $i$ on this grid has a local energy density $u_i$, polarization $P_i$ and internal loss $\rho_i$. In accordance with Eq. (\ref{eq1}), the local energy density at point $i$ of the ferroelectric is then given by 

\begin{equation}
u_{i} = \alpha_{i} P_{i}^2 + \beta_{i} P_{i}^4 + \gamma_{i} P_{i}^6 - EP_{i} + k\sum_{j}\left(\frac{P_i-P_j}{\Delta x}\right)^2,
\label{eq4}
\end{equation}

where $\alpha_i$, $\beta_i$ and $\gamma_i$ are the local anisotropy constants, $\Delta x$ is the grid spacing and the index $j$ denotes all cells directly adjacent to cell $i$. While cells in the middle of the grid have four nearest neighbors (left, right, above and below), cells at the edge or at the corner of the grid have three and two nearest neighbors, respectively, corresponding to open boundary conditions at the capacitor edge. To calculate the total charge on the ferroelectric capacitor $Q_F$, the individual contributions of each cell have to be summed up as 

\begin{equation}
Q_F = A \left( \epsilon_0 E + \frac{1}{N}\sum_{i=1}^N P_i\right) \approx \frac{A}{N}\sum_{i=1}^N P_i .
\label{eq6}
\end{equation}

Here, $A$ is the total area of the ferroelectric capacitor, $\epsilon_0$ is the vacuum permittivity and $N$ is the total number of cells. The approximation in Eq. (\ref{eq6}) is reasonable, because $P >> \epsilon_0 E$ in almost all ferroelectrics. Since $\Delta x = \Delta y$, the area of each cell is given by $A/N = (\Delta x)^2$ and therefore $\Delta x = \sqrt{A/N}$.

The effective field $E_{eff}$ in the ferroelectric is given by 

\begin{equation}
E_{eff} = \frac{V_F}{t_F} - E_{bias},
\label{eq7}
\end{equation}

where $V_F$ is the voltage across the ferroelectric capacitor, $t_F$ is the thickness of the ferroelectric and $E_{bias}$ is an internal bias field, which might be induced by electrodes with different work functions or inhomogeneous distribution of charges along the $z$-axis in the ferroelectric.\cite{schenk_complex_2015,pesic_nonvolatile_2016,pesic_physical_2016} The connection to the external circuit as seen in Fig. \ref{fig1} is now easily defined by Kirchhoff's voltage and current laws. The voltage defined by the pulse generator $V_S$ can also be written as 

\begin{equation}
V_S = i_R R + V_F,
\label{eq8}
\end{equation}

where $i_R$ is the current flowing through the external resistor $R$. Since there is also a parasitic capacitance $C_p$ related to the measurement setup in parallel to the ferroelectric, $i_R$ can be expressed as 

\begin{equation}
i_R = i_F + i_{Cp} = \frac{\partial Q_F}{\partial t} + C_p \frac{\partial V_F}{\partial t}.
\label{eq9}
\end{equation}

Here, $i_F$ is the total ferroelectric current and $i_{Cp}$ is the current through the parasitic capacitor $C_p$. The complete dynamic behavior of the circuit in the diffusive limit approximation of the ferroelectric is described by Eqs. (\ref{eq2}-\ref{eq9}). Leakage currents through the ferroelectric are neglected here, because their effect on NC voltage transients was shown to be minimal as long as they are much lower compared to $i_R$.\cite{hoffmann_direct_2016}

To account for any non-uniformity in the ferroelectric layer, spatial distributions of the parameters $\alpha$, $\beta$, $\gamma$, $\rho$ and $E_{bias}$ can be defined. This approach can emulate different nucleation times for reverse domain formation as well as general defect driven property variations in the ferroelectric.

\section{Results and Discussion}

\begin{figure*}
\includegraphics[width=17.8cm]{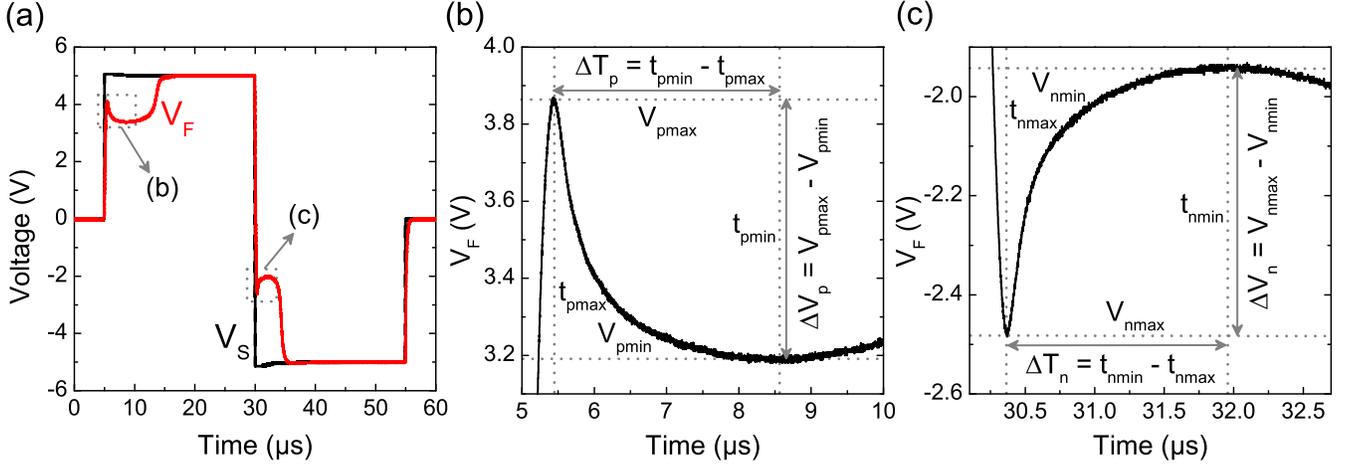}
\caption{(a) Voltage transients measured across the pulse generator $V_S$ and the ferroelectric $V_F$ for a capacitor with $A = (50 \ \mu\text{m})^2$ and $R = 3360 \ \Omega$. Magnifications of the time frames where the ferroelectric capacitance is negative are shown in (b) for positive $V_S$ and in (c) negative $V_S$.} \label{fig2}
\end{figure*}

In this section, first, typical measured NC voltage transients are presented to define characteristic NC parameters directly obtainable from the experiment. Time-dependent Ginzburg-Landau simulations will then be fitted to this experimental data, to gain insights into the domain switching dynamics during the NC transient. Subsequently, the effect of experimental measurement conditions on the NC transients will be investigated. Initially, the influence of applied and internal electric fields in the ferroelectric will be discussed. Afterwards, the effect of the capacitor area and the external resistor on the NC transients are examined. Furthermore, it will be shown that a simple analytical model can be developed to extract intrinsic material parameters from these measurements, namely the average NC per area, the activation field and the average internal loss.

\subsection{Characterizing Ferroelectric Negative Capacitance Voltage Transients}

An exemplary NC voltage transient of a PZT capacitor with $A = (50 \ \mu \text{m})^2$, an external resistance $R = 3360 \ \Omega$ and a maximum voltage amplitude $V_{S,max} = 5$ V is shown in Fig. \ref{fig2}. Initially, before the pulse $V_S$ shown in Fig. \ref{fig2}(a) is applied, the ferroelectric capacitor is completely poled into a state where the spontaneous polarization vector points from the bottom towards the top electrode. Therefore, when applying a positive voltage pulse $V_S$, the voltage $V_F$ initially increases, until at a certain voltage $V_F = V_{pmax}$ the polarization will begin to switch. A magnification of this time frame is shown in Fig. \ref{fig2}(b). During switching, for a certain time $t_{pmax} < t < t_{pmin}$, the voltage across the ferroelectric decreases. As was pointed out by Khan \textit{et al}., this behavior can be interpreted as the ferroelectric possessing a negative differential capacitance, since $V_F$ decreases while $Q_F$ increases and the ferroelectric capacitance, defined as $C_F = \text{d}Q_F/\text{d}V_F$, is negative in this case.\cite{khan_negative_2014}

\begin{figure}
\includegraphics[width=7.5cm]{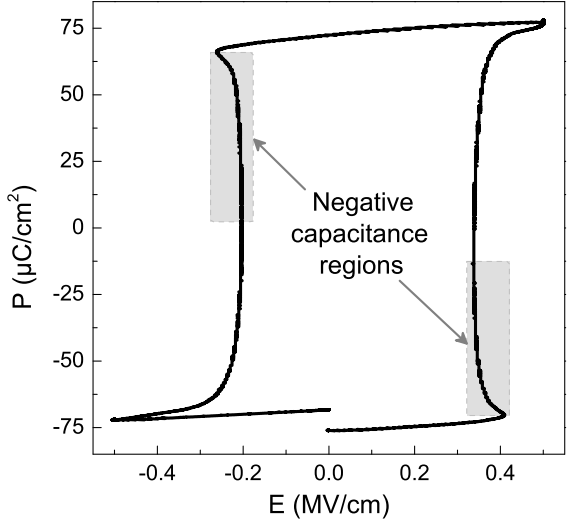}
\caption{Polarization-electric field hysteresis for a capacitor with $A = (50 \ \mu\text{m})^2$ and $R = 3360 \ \Omega$ calculated from negative capacitance voltage transients shown in Fig. \ref{fig2}.} \label{fig3}
\end{figure}

At $t = t_{pmin}$, the voltage $V_F$ shows a minimum $V_{pmin}$ after which $V_F$ again increases with time and therefore $C_F > 0$. After the switching and charging of the ferroelectric is completed, no current $i_R$ is flowing anymore and $V_S = V_F$. However, when the applied voltage $V_S$ is reversed to -5 V, a similar NC voltage transient can be observed during the back-switching process. A zoom-in on this second time frame $t_{nmax} < t < t_{nmin}$ where $C_F < 0$ is shown in Fig. \ref{fig2}(c). Analogous to the previous case, the NC transient begins at $V_F = V_{nmax}$ and terminates at $V_F = V_{nmin}$. Such transient behavior during switching in BaTiO$_3$ was actually already experimentally reported by Merz in 1954.\cite{merz_domain_1954} However, it was not interpreted as a transient NC at that time. Furthermore, if $V_S(t)$, $V_F(t)$, $R$ and $C_p$ are known, the charge of the ferroelectric can be calculated as \cite{hoffmann_direct_2016}

\begin{equation}
Q_F = Q_{F0} + \frac{1}{R} \int \limits_0^t \left[V_S(t')-V_F(t') \right]\text{d}t'-V_F C_p.
\label{eq10}
\end{equation}

\begin{figure*}
\includegraphics[width=18.4cm]{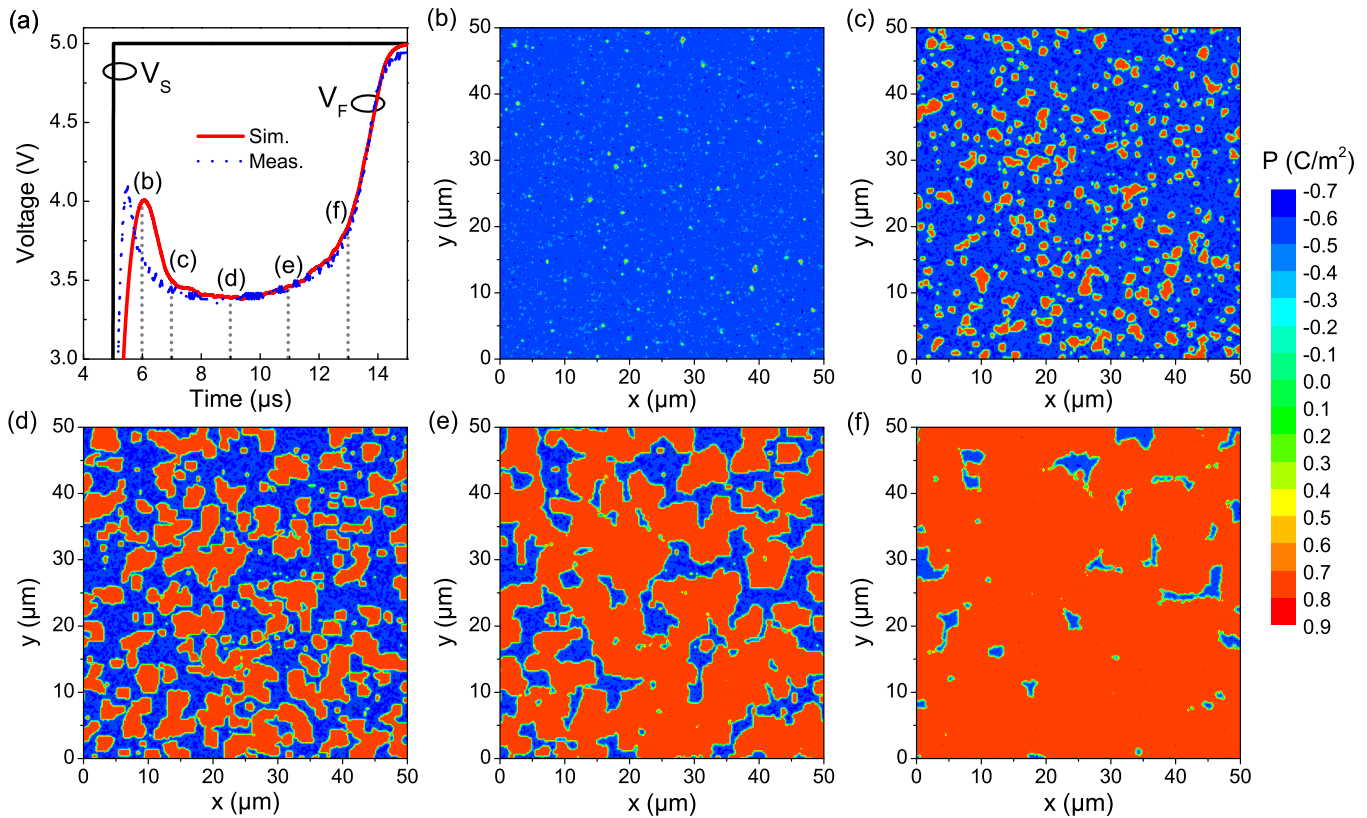}
\caption{(a) Simulated voltage transients across the ferroelectric $V_F$ in response to the applied voltage $V_S$. (b)-(f) Spatial polarization distribution in the ferroelectric during switching for t = 6 $\mu$s, 7 $\mu$s, 9 $\mu$s, 11 $\mu$s and 13 $\mu$s, respectively.} \label{fig4}
\end{figure*}

Here, $Q_{F0} = Q_F(t=0)$ is the initial charge of the ferroelectric. From Eq. (\ref{eq10}), we can now calculate the polarization $P \approx Q_F/A$ and plot the $P$-$E$ hysteresis of the ferroelectric, which is shown in Fig. \ref{fig3}. The loop is not completely closed due to the presence of leakage currents for positive applied fields. The NC regions during switching can be directly seen from the gray shaded areas in Fig. \ref{fig3}, where d$P$/d$E < 0$. These regions are predicted by the S-shaped $P$-$E$ relationship which is implicitly defined by Landau-Devonshire theory and was already predicted in several studies over a decade ago.\cite{ricinschi_analysis_1998,bratkovsky_depolarizing_2006} However, it is clear that there is still hysteresis, because the NC regions are not stabilized in this isolated capacitor configuration and thus the NC behavior is only transient in nature. To stabilize NC, a ferroelectric-dielectric bilayer structure as reported before is necessary,\cite{islam_khan_experimental_2011,zubko_negative_2016} otherwise this effect will always be hysteretic because of spinodal decomposition into a multi-domain state.\cite{artemev_spinodal_2010} The relationship between domain formation and hysteresis will become more apparent through the simulation results in the following section.

\subsection{Simulating Negative Capacitance Domain Dynamics}

To understand the domain dynamics during the NC voltage transients in more detail, time-dependent Ginzburg-Landau simulations were carried out and simulation parameters ($\alpha$, $\beta$, $\gamma$, $\rho$, $E_{bias}$ and $k$) were fitted to the experimental data shown in Fig. \ref{fig2}. A good agreement was achieved by using the parameter distributions listed in Table \ref{tab1}. For $\alpha$, $\beta$, $\gamma$, and $E_{bias}$, Gaussian distributions with a standard deviation $\sigma$ of 7.5 \% with respect to the mean values $\mu$ were used. The internal loss was assumed to be uniformly distributed with the minimum being $\rho_{min} = 0.4 \ \Omega$m and a maximum of $\rho_{max} = 13 \ \Omega$m. The domain wall coupling constant $k$ was fitted to be $k = 2.5 \times 10^{-7}$ m\textsuperscript{3}/F.

\begin{table}
\caption{Parameters fitted from the experimental data, which were used for time-dependent Ginzburg-Landau simulations. $\alpha$, $\beta$, $\gamma$ and $E_{bias}$ are normally distributed, $\rho$ is uniformly distributed.}
\begin{ruledtabular}
\begin{tabular}{cccc}
& \textbf{$\mu$} & \textbf{$\sigma$} & \textbf{unit} \\
$\alpha$ & $-4.4 \times 10^7$ & $-3.3 \times 10^6$ & m/F\\
$\beta$ & $-9.1 \times 10^6$ & $-6.8 \times 10^5$ & m\textsuperscript{5}/(C\textsuperscript{2}F) \\
$\gamma$ & $7.4 \times 10^7$ & $5.6 \times 10^6$ & m\textsuperscript{9}/(C\textsuperscript{4}F) \\
$E_{bias}$ & $7.2 \times 10^6$ & $5.4 \times 10^5$ & V/m \\
\hline
& \textbf{value} & \textbf{unit} & \\
$k$ & $2.5 \times 10^{-7} $ & m$^3$/F & \\
$\rho_{min}$ & 0.4 & $\Omega$m & \\
$\rho_{max}$ & 13 & $\Omega$m & \\
\end{tabular}
\end{ruledtabular}
\label{tab1}
\end{table}

The resulting simulated NC voltage transient for $V_S > 0$ is shown in Fig. \ref{fig4}(a). It can be seen that a very similar transient NC behavior can be observed when compared to the experimental data from Fig. \ref{fig2}(b). Only during the initial charging and switching ($t < 7 \ \mu$s) of the capacitor, there are some deviations between simulation and measurement: The change of $V_F$ with time is slightly steeper and the maximum of $V_F$ appears earlier in the measurement compared to the simulation, which shows that the initial switching actually happens slightly faster than our model was capable of reproducing. The reason for this is the diffusive limit approximation of Eq. (\ref{eq2}), which cannot quantitatively predict the intrinsic dynamics during initial switching.

For $t > 7 \ \mu$s on the other hand, excellent agreement between data and theory was achieved, since here the switching dynamics can be accurately described by an effective domain wall velocity. Furthermore, in Fig. \ref{fig4}(b)-(f) the spatial polarization distribution across the capacitor area is shown as contour plots for different points in time during switching. At the beginning of the switching process ($t = 6 \ \mu$s), Fig. \ref{fig4}(b) shows that first reverse domains nucleate randomly distributed across the capacitor area. Going further in time and looking at Fig. \ref{fig4}(c), the effective capacitance of the ferroelectric is still negative and some indication of the origin can be extracted from the spatial polarization distribution: While some new reverse nucleation sites are still forming, already nucleated domains can grow sideways almost unimpeded. However, the negative slope of d$V_F$/d$t$ is beginning to flatten out. This can be understood from the shrinking area that is available for reverse domain nucleation and unrestricted growth which is known from Kolmogorov-Avrami-Ishibashi (KAI) switching dynamics in epitaxial PZT thin films and single crystals.\cite{ishibashi_note_1971} Therefore, at a certain point in time ($t = 9 \ \mu$s), $V_F$ will stop to decrease and start increasing with time again. From the contour plot in Fig. \ref{fig4}(d) it can be seen that at this point in time, about half of the capacitor area is already switched and unrestricted domain wall motion is mostly replaced by domain coalescence. In Fig. \ref{fig4}(e) and (f), more domains coalesce while less area of the ferroelectric capacitor is switching which results in an increase of $V_F$ with time, which is expected for charging of a positive capacitor. 

This shows, that in case of $A = (50 \ \mu\text{m})^2$ and $R = 3360 \ \Omega$, the NC transient times $\Delta T_{p,n}$ in these epitaxial PZT films are mostly limited by the domain wall velocity and not by the initial domain nucleation. However, most of the change of $V_F$ during the initial NC transient seems to happen during this initial reverse nucleation process. Very similar reverse domain nucleation and growth processes in Pb(Zr$_{0.2}$Ti$_{0.8}$)O$_3$ were shown to also behave according to Merz's law by atomic force microscopy experiments.\cite{paruch_nanoscale_2006} From these results and also visible in Fig. \ref{fig4}, it is apparent that such transient NC effects will always be hysteretic due to the underlying irreversible switching processes. For example, when the ferroelectric is in the NC state, see Fig. \ref{fig4}(b)-(d), applying a large negative voltage will not continue the NC effect in the opposite direction, because there is already coalescence of the -$P_s$ (blue) domains happening in this state, thus the capacitance will be positive. Therefore, the whole ferroelectric has to be switched to the opposite polarization state before another NC transient can be measured. This will always result in a large $P$-$E$ hysteresis in such capacitors. This also implies, that transient NC cannot be used for small-signal voltage amplification. Therefore, transient NC effects in ferroelectric capacitors have rather different underlying physics compared to stabilized NC in a ferroelectric-dielectric bilayer or superlattice, where the stabilization of a small-signal NC was shown to be possible. \cite{zubko_negative_2016} However, also this stabilized NC seems to stem from a complex domain structure resulting in a negative contribution to the permittivity due to domain wall motion.

In a next step, the influence of the electric field in the ferroelectric on the NC switching times $\Delta T_p$ and $\Delta T_n$, will be investigated. 

\subsection{Influence of the Applied Voltage}

Since the NC effect in epitaxial PZT seems to be a result of domain wall motion, the effect of different applied voltages $V_S$ was investigated. Therefore, capacitors with $A = (35 \ \mu \text{m})^2$ were connected to a series resistor $R = 10$ k$\Omega$ and the amplitude of $V_S$ was changed, while the duration of the voltage pulses was kept constant. It is intuitively expected, that by applying a larger voltage pulse $V_S$ to the series connection, the NC transient should get shorter, because then more charge can be supplied to the capacitor in the same amount of time.

\begin{figure}
\includegraphics[width=7.5cm]{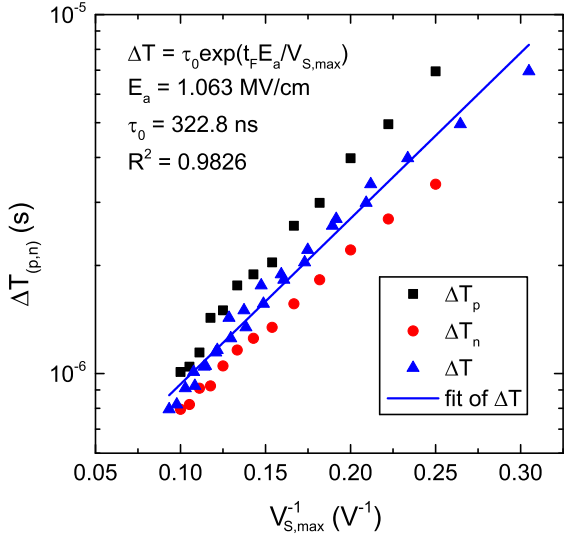}
\caption{Negative capacitance transient times with ($\Delta T$) and without correction ($\Delta T_{p,n}$) for internal bias fields as a function of $V_{S,max}$ with $R = 10$ k$\Omega$ and $A = (35 \ \mu \text{m})^2$. Solid line shows analytical fit of $\Delta T$.} \label{fig8}
\end{figure}

In Fig. \ref{fig8}, the characteristic NC transient times $\Delta T_{p,n}$ are plotted against $1/V_{S,max}$, which is the reciprocal of the maximum applied voltage $V_S$. The reason for this is the very linear relationship between $\Delta T_{p,n}$ and $1/V_{S,max}$ in this semi-logarithmic representation in Fig. \ref{fig8}. This apparent linearity is related to the electric field dependence of the domain wall velocity known empirically as Merz's law:\cite{merz_domain_1954}

\begin{equation}
\frac{1}{v_{DW}} \propto \tau_{sw} = \tau_0 \text{exp}\left(\frac{E_a}{E} \right)^m.
\label{eq11}
\end{equation}

Here, $v_{DW}$ is the domain wall velocity, $\tau_{sw}$ is the characteristic switching time, $\tau_0$ is the intrinsic switching time constant, $E_a$ is the activation field and $m$ is the exponent. In the initial work of Miller and Weinreich, this dependence was attributed to the nucleation of triangular steps at the domain wall.\cite{miller_mechanism_1960} However, this approach yielded activation fields an order of magnitude larger than experimentally observed. \cite{meng_velocity_2015} In the case where $m = 1$, the domain wall motion is a creep process,\cite{tybell_domain_2002} which is ascribed to random field defects.\cite{liu_intrinsic_2016} From the activation field $E_a$ in Eq. (\ref{eq11}), the domain wall energy density can be calculated when neglecting the depolarization energy of the reverse domain nucleus:\cite{hu_universal_2014}

\begin{equation}
E_a = \frac{c\sigma_{DW}^2}{P_s k_B T} .
\label{eq12}
\end{equation}

Here, $c$ is the domain wall width, which can be approximated as the in plane lattice constant, $\sigma_{DW}$ is the domain wall energy density and $P_s$ is the spontaneous polarization. Once we have obtained $E_a$ from the experimental data, we will estimate $\sigma_{DW}$ from Eq. (\ref{eq12}).

However, when looking at Fig. \ref{fig8}, we can see that there are two different trends for positive and negative NC transients, $\Delta T_p$ and $\Delta T_n$, respectively. Since it was already established that our PZT capacitors exhibit an internal bias field ($t_F E_{bias} = 0.72$ V), it makes sense to correct this offset by plotting $\Delta T_p$ against 1/$(V_{S,max}-t_F E_{bias})$ and $\Delta T_p$ against 1/$(V_{S,max}+t_F E_{bias})$ as also shown in Fig. \ref{fig8}. As can be seen, after correcting $\Delta T_p$ and $\Delta T_n$ for $E_{bias}$, all data points now lie on the same trend line and are therefore labeled only $\Delta T$.

Now, by fitting Eq. (\ref{eq11}) to the experimental data ($\Delta T$ vs. $V_{S,max}^{-1}$) shown in Fig. \ref{fig8}, we can estimate $E_a$ and therefore also $\sigma_{DW}$. In this case, we use the duration of the NC transient $\Delta T$ as the characteristic switching time $\tau_{sw}$ with $m = 1$ and get $\tau_0 = 322.8$ ns and $E_{a} = 1.063$ MV/cm. This value for the activation field $E_a$ is in very good agreement with the results reported before ($E_a \approx 1$ MV/cm).\cite{tybell_domain_2002,shin_nucleation_2007} Using Eq. (\ref{eq12}) with $c = 3.95$ {\AA} \cite{frantti_neutron_2000} and $P_s = 0.7$ C/m$^2$, we can then calculate the domain wall energy density as $\sigma_{DW} = 27.93$ mJ/m$^2$. This value is rather low compared to \textit{ab initio} calculations for PbTiO$_3$ ($\sigma_{DW} = 132$ mJ/m$^2$)\cite{meyer_ab_2002} and experimental values for Pb(Zr$_{0.2}$Ti$_{0.8}$)O$_3$ ($\sigma_{DW} = 120$ mJ/m$^2$).\cite{ganpule_polarization_2001} The reason for this discrepancy might be a domain wall width $c$, which is larger than the lattice constant or an underestimation due to the neglect of the depolarization energy in the derivation of Eq. (\ref{eq12}).\cite{hu_universal_2014} Additionally, \textit{ab initio} calculations were carried out only for $T = 0$ K.\cite{meyer_ab_2002}

Nevertheless, the field independent switching time $\tau_0 = 322.8$ ns, fitted to the data in Fig. \ref{fig8}, might also give us an insight into the intrinsic material parameters of the sample, like the average loss $\rho$ and the average NC per area $C_{FE}$. For this, we will need an analytical model and vary the circuit parameters like capacitor area $A$ and external resistance $R$. Since the NC transient duration $\Delta T$ is clearly limited by a domain wall creep process, we will investigate how the domain wall velocity can be precisely controlled by limiting the charge supply through the external circuit. 

\subsection{An Analytical Model for Transient Negative Capacitance}

From Eq. (\ref{eq11}) we already know the field dependence of $\Delta T$ and also the temperature dependence from $E_a = f(T)$ in Eq. (\ref{eq12}). While this field and temperature dependence comes from the domain wall dynamics of the ferroelectric itself, the external circuit will also strongly influence the transient NC effect as reported before.\cite{khan_negative_2014,hoffmann_direct_2016} Therefore, a simplified electrical circuit as shown in Fig. \ref{fig1} is considered, where the ferroelectric can be modeled according to the Landau-Khalatnikov equation (\ref{eq2}) as a non-linear capacitor in series with a resistance $\rho t_F/A$, which describes the loss in ferroelectric. Since we are now only interested in the influence of the external circuitry, we neglect the Ginzburg term $k(\nabla P)^2$ as well as any parasitic capacitance or resistances. Then we can write Kirchhoff's voltage law

\begin{equation}
V_S \approx \frac{\text{d} Q_F}{\text{d} t} \left(R  + t_F\rho/A\right) + \left(2 \alpha P + 4 \beta P^3 + 6\gamma P^5\right)t_F,
\label{eq13}
\end{equation}

where $\text{d} Q_F / \text{d} t$ is the current flowing through $R$ and $\rho t_F/A$ and $Q_F \approx P A$ is the charge on the ferroelectric capacitor. Therefore, we can write

\begin{equation}
V_S \approx \frac{\text{d} P}{\text{d} t} \left(RA  + t_F\rho\right) + \left(2 \alpha P + 4 \beta P^3 + 6\gamma P^5\right)t_F.
\label{eq14}
\end{equation}

Since we are interested in the change of $P$ with time, we can transform Eq. (\ref{eq14}) and get

\begin{equation}
\frac{\text{d} P}{\text{d} t} \approx \frac{V_S - \left(2\alpha P + 4 \beta P^3 + 6\gamma P^5\right) t_F}{RA + t_F\rho}.
\label{eq15}
\end{equation}

From Eq. (\ref{eq15}) it is now apparent that the time d$t$ necessary to change the polarization by d$P$ is proportional to $(RA + t_F\rho)$. This shows that there is an intrinsic term $t_F\rho$ and an extrinsic term $RA$ limiting the switching speed of the ferroelectric. The limited charge supply through the external resistor might also help to explain the strong dependence of the domain wall velocity on the experimental measurement time, reported by other authors.\cite{meng_velocity_2015} This also shows that there are two different NC switching regimes: First, if $RA << t_F \rho$, the polarization reversal is limited by the internal loss in the ferroelectric material, and changing either $R$ or $A$ will not significantly increase the switching time. This can be seen in the case of NC transient measurements on ferroelectric doped HfO$_2$ capacitors for different $R$.\cite{hoffmann_direct_2016} In the other case, $RA >> t_F \rho$ and the internal loss of the ferroelectric can be neglected compared to the limiting external circuit. This is the case for our simulation and measurement results shown in Fig. \ref{fig2} and Fig. \ref{fig8}. In this regime, the NC transient time should be proportional to $R$ and $A$. Therefore, we can now propose an analytical model for the NC transient time $\Delta T$ based on Eq. (\ref{eq11}) and Eq. (\ref{eq15}):

\begin{equation}
\Delta T = |C_{FE}|(RA + t_F\rho)\text{exp}\left(\frac{E_a}{E} \right).
\label{eq16}
\end{equation}

There are three free parameters in this model: the average NC per area $C_{FE}$, the average internal loss $\rho$ and the activation field $E_a$, which we have already determined from field-dependent NC transient measurements. To determine $C_{FE}$ and $\rho$, further measurements on differently sized capacitors (changing $A$) or different external resistors $R$ are suitable. Therefore, in the following section, we investigate the effect of the capacitor area on the NC voltage transients.

\subsection{Influence of the Capacitor Area}

PZT capacitors with different areas from \mbox{(10 $\mu$m)\textsuperscript{2}} to \mbox{(50 $\mu$m)\textsuperscript{2}} were fabricated and characterized in the same way as was shown for \mbox{$A = (50 \ \mu$m)\textsuperscript{2}} in Fig. \ref{fig2}. The external resistance was kept constant at $R = 10 \ \text{k}\Omega$ and the maximum applied voltage was $V_{S,max} = 5$ V.

With the knowledge of the internal bias and activation field $E_a$ from the field-dependent measurement results, we can also correct the area dependent data for this internal bias of $t_F E_{bias} = 0.72$ V. Therefore, we can use the equation

\begin{equation}
\Delta T = \Delta T_{p,n} \text{exp} \left[E_a \left(\frac{1}{E}-\frac{1}{E\pm E_{bias}} \right) \right]
\label{eq17}
\end{equation}

\begin{figure}
\includegraphics[width=7cm]{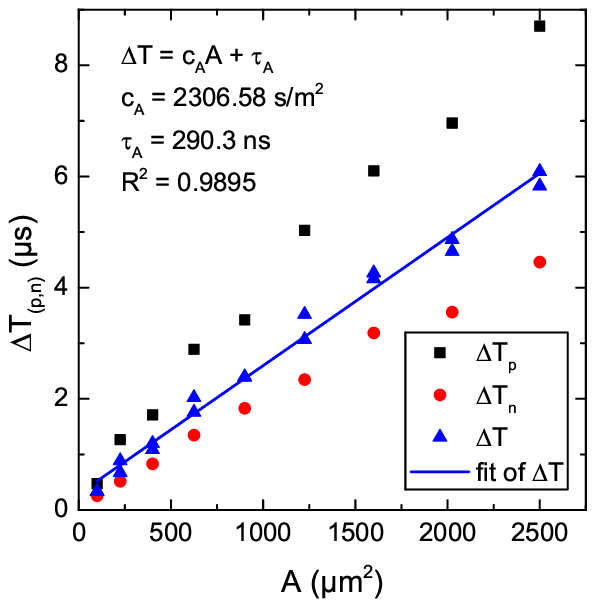}
\caption{Negative capacitance transient times with ($\Delta T$) and without correction ($\Delta T_{p,n}$) for internal bias fields as a function of $A$ with $R = 10$ k$\Omega$ and $V_{S,max} = 5$ V. Solid line shows linear fit of $\Delta T$.} \label{fig12}
\end{figure}

to calculate the actual NC transient time $\Delta T$ from $\Delta T_p$ and $\Delta T_n$ which are different due to the broken symmetry because of the internal bias field. In Eq. (\ref{eq17}), $E- E_{bias}$ and $E+ E_{bias}$ is used when calculating $\Delta T_n$ and $\Delta T_p$, respectively. The resulting dependence of $\Delta T$ and for comparison $\Delta T_{p,n}$ on $A$ is shown in Fig. \ref{fig12}. As expected, after correcting $\Delta T_{p,n}$ for $E_{bias}$, only a single linear trend for $\Delta T$ vs. $A$ is observed. A linear fit ($\Delta T = c_A A + \tau_A$) to the data is also presented in Fig. \ref{fig12}. With these fitted values $c_A$ and $\tau_A$, the parameters $C_{FE}$ and $\rho$ of our analytic model in Eq. (\ref{eq16}) can be determined from

\begin{equation}
\begin{split}
c_A &= R|C_{FE}|\text{exp}\left(\frac{E_a}{E}\right), \\
\tau_A &= t_F\rho |C_{FE}|\text{exp}\left(\frac{E_a}{E}\right).
\end{split}
\label{eq18}
\end{equation}

From Eq. (\ref{eq18}) we can now calculate the average NC per unit area as 

\begin{equation}
|C_{FE}| = \frac{c_A}{R}\text{exp}\left(-\frac{E_a}{E}\right).
\label{eq19}
\end{equation}

By using Eq. (\ref{eq19}) with $c_A$ and $E_a$ fitted from the experimental data in Fig. \ref{fig8} and Fig. \ref{fig12}, respectively, we can determine $C_{FE} = -0.023$ F/m$^2$. It should be noted here, that this value is about 5 times lower than the theoretical value $C_{FE} \approx 1/(2\alpha t_F)$ which can be estimated based on homogeneous Landau theory (in our case -0.114 F/m$^2$), which is almost always used in the literature when calculating the stability condition of NC in series with a positive capacitor. Looking at our analytical model in Eq. (\ref{eq16}), we can see that $C_{FE}$ is actually just the NC in the limit where $E \rightarrow \infty$ and the NC that is actually measured can be expressed as $C_{FE,m} = C_{FE}\text{exp}(E_a/E)$. Therefore, $C_{FE,m}$ seems to be proportional to the domain wall velocity $v_{DW}$, while $C_{FE}$ is an intrinsic material parameter which is independent of $E$. This is actually in contrast to a rather voltage-independent NC which was extracted in a previous study.\cite{khan_negative_2014} However, the determination of $C_{FE}$ shown here, seems to be much more reliable.

We now can also make a first estimate of the average of $\rho$ using Eq. (\ref{eq18}), with the newly calculated $C_{FE}$ and $\tau_A$ from Fig. \ref{fig12} and we get $\rho \approx 15 \ \Omega$m. However, this results should be treated with caution, since the fit in Fig. \ref{fig12} might not be very accurate for $A \rightarrow 0$. Therefore, a more reliable approach is to vary $R$, while keeping $A$ and $V_S$ constant, which will be elucidated in the following.

\subsection{Influence of the Series Resistance}

In this section, the effect of a change of $R$ on the NC voltage transients will be investigated for a constant capacitor area of 2500 $\mu$m\textsuperscript{2} and $V_{S,max} = 5$ V. As expected and also shown in previous studies,\cite{khan_negative_2014,hoffmann_direct_2016} a larger $R$ will slow down the switching as can be seen from Eq. (\ref{eq15}). When $R$ increases, d$P$/d$t$ is reduced and thus also the NC transient time increases as shown in the analytical Eq. (\ref{eq16}).

\begin{figure}
\includegraphics[width=7.5cm]{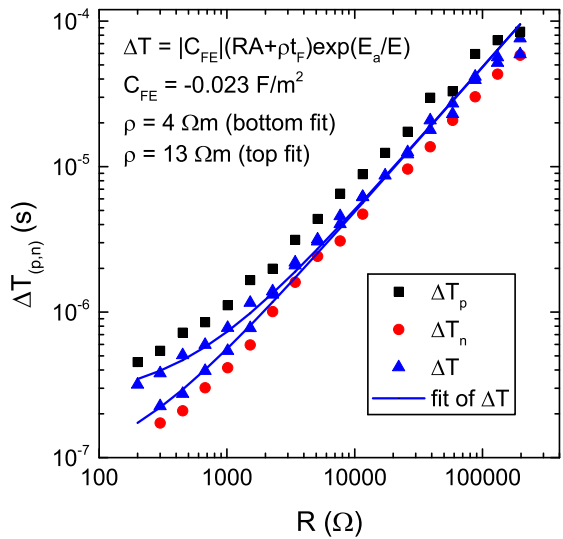}
\caption{Negative capacitance transient times with ($\Delta T$) and without correction ($\Delta T_{p,n}$) for internal bias fields as a function of $R$ with $A = (50 \ \mu \text{m})^2$ and $V_{S,max} = 5$ V. Solid lines show linear fit of $\Delta T$.} \label{fig16}
\end{figure}

To validate our newly developed analytical model for the NC transient time $\Delta T$ in Eq. (\ref{eq16}) and to determine the average value of $\rho$ more precisely than our previous estimation of 15 $\Omega$m, first, $\Delta T$ has to be calculated from Eq. (\ref{eq17}) to correct $\Delta T_{p,n}$ for the internal bias field. The dependence of $\Delta T$ on $R$ as well as $\Delta T_{p,n}$ for comparison is shown in Fig. \ref{fig16}. While, $E_a = 1.063$ MV/cm and $C_{FE} = -0.023$ F/m$^2$ where used from the fitted experimental data in Fig. \ref{fig8} and Fig. \ref{fig12}, respectively, $\rho$ was chosen to fit the $R$ dependent data in Fig. \ref{fig16}. From this figure it is apparent that $\rho$ is different for the positive and negative switching transients even after internal bias correction for small values of $R$: The NC transient times for $V_S < 0$ are slightly shorter compared to the ones $V_S > 0$, which means that $\rho$ is larger in the latter case. From Fig. \ref{fig16} we can see an excellent agreement of our analytical model to the experiment, when using $\rho = 4 \ \Omega$m for the $V_S < 0$ transient and $\rho = 13 \ \Omega$m for the $V_S > 0$ transient. Interestingly, these values are very similar to the range of randomly distributed $\rho$ values fitted for the time-dependent Ginzburg-Landau simulations, see Table \ref{tab1}.

Finally, we can again validate our analytical model by calculating $C_{FE}$ and $\rho$ from the fit to the field-dependent data in Fig. \ref{fig8} by taking either one of the two from the previous fit in Fig. \ref{fig16}. Using $C_{FE} = -0.023$ F/m$^2$ we then get $\rho = (\tau_0/|C_{FE}|-RA)/t_F = 17.8 \ \Omega$m. On the other hand, when using $\rho = 4 \ \Omega$m to 13 $\Omega$m, we can calculate $C_{FE} = -\tau_0/(RA+t_F\rho) = -0.026$ F/m$^2$ and $C_{FE} = -0.024$ F/m$^2$, respectively. These results are very similar to the ones estimated before, showing the consistency of our proposed model described by Eq. (\ref{eq16}).

\section{Summary and Outlook}

We have investigated NC voltage transients during polarization switching in high quality epitaxial capacitors of ferroelectric Pb(Zr$_{0.2}$Ti$_{0.8}$)O$_3$ connected in series with an external resistor. A time-dependent Ginzburg-Landau approach was used to gain insight into the transient switching behavior and domain dynamics using numerical simulations. Three main experimental parameter variations are investigated: A change of the applied electric field $E$, the capacitor area $A$ and the external resistance $R$. In particular, the duration of the NC voltage transients during switching from positive to negative polarization and vice versa were analyzed. The simulation results indicate, that in the regime where the NC transient time is limited by transverse domain wall motion and coalescence, the NC effect is mostly caused by reverse domain nucleation and unrestricted growth. With the onset of domain coalescence, the differential capacitance of the ferroelectric capacitor becomes positive again. 

By investigating the electric field dependence of the NC transient time, it was shown that the transverse domain wall velocity is the decisive factor limiting the switching speed. This was corroborated by the observation of a typical domain wall creep behavior in accordance with Merz's law. Furthermore, it was shown that the different NC durations when applying a positive or negative voltage pulse can be attributed to internal bias fields. By correcting for this internal field, an activation field of $\sim$1 MV/cm was determined which in turn allowed for an estimation of the domain wall energy density as $\sim$28 mJ/m$^2$. An analytical model was developed to relate the NC transient time $\Delta T$ to the experimental parameters $R$, $A$ and $E$ which allows to reliably determine the average intrinsic NC per area $C_{FE}$ and the average internal loss $\rho$ from measurement data. While $\Delta T$ is proportional to both $R$ and $A$, the field dependence is exponential. For 100 nm thin films we obtained $C_{FE}= -0.023$ F/m$^2$ and $\rho = 4 \ \Omega$m and $13 \ \Omega$m for negative and positive applied fields, respectively. 

This study shows that NC voltage transients in epitaxial ferroelectric capacitors can be explained by reserve domain nucleation and growth dynamics which can be accurately modeled by time-dependent Ginzburg-Landau theory. Since these processes are known to be irreversible and dissipative, there will always be hysteresis effects associated with NC in isolated ferroelectric capacitors. This has to be considered when considering device applications like low power logic transistors, where hysteresis should be avoided. For this case, integration of the ferroelectric directly into the gate stack seems mandatory.

\section*{Acknowledgments}

M.H., M.P., U.S., S. Slesazeck, and T.M. gratefully acknowledge the support
by the European Fund for Regional Development and the Free State of
Saxony. M.H. acknowledges a visiting scholar fellowship from the University of California, Berkeley. M.H. also acknowledges fruitful discussions with Samuel Smith at UC Berkeley.

\end{document}